\documentclass[twocolumn]{aastex63}
\usepackage{amsmath}
\usepackage{booktabs}
\usepackage{multirow}
\usepackage{enumitem}

\newcommand{\msun}{${\rm M_{\sun}}$}
\def\ltsima{$\; \buildrel < \over \sim \;$}
\def\simlt{\lower.5ex\hbox{\ltsima}}
\def\gtsima{$\; \buildrel > \over \sim \;$}
\def\simgt{\lower.5ex\hbox{\gtsima}}
\def\km{{\rm\,km}}
\def\kms{{\rm\,km\,s^{-1}}}
\def\km2s2{{\rm\,km^2\,s^{-2}}}

\def\kpc{{\rm\,kpc}}

\def\msun{{\rm\,M_\odot}}


\def\Gyr{{\rm\,Gyr}}

\def\ltsima{$\; \buildrel < \over \sim \;$}
\def\gtsima{$\; \buildrel > \over \sim \;$}

\def\ins{{\it in-situ}}
\def\accreted{{\it accreted}}
\def\Splash{{\it Splash}}
\def\Aurora{{\it Aurora}}
\def\Sagittarius{{\it Sagittarius}}
\def\Gaia{{\it Gaia}}
\def\GSE{{\it GSE}}
\def\Shakti{{\it Shakti}}
\def\Shiva{{\it Shiva}}
\def\POH{{\it POH}}
\def\LMC{{\it LMC}}
\def\SMC{{\it SMC}}

\submitjournal{ApJ}

\shorttitle{Two new substructures in the Milky Way}
\shortauthors{Malhan \& Rix}
\begin{document}

\title{SHIVA and SHAKTI:\\Presumed Proto-Galactic Fragments in the Inner Milky Way}

\correspondingauthor{Khyati Malhan}
\email{kmalhan07@gmail.com}

\author[0000-0002-8318-433X]{Khyati Malhan}
\affiliation{Humboldt Fellow and IAU Gruber Fellow}
\affiliation{Max-Planck-Institut f\"ur Astronomie, K\"onigstuhl 17, D-69117, Heidelberg, Germany}
\affiliation{Myrspoven AB, V\"astg\"otagatan 1, 11827 Stockholm, Sweden}
%\nocollaboration{1}

\author[0000-0003-4996-9069]{Hans-Walter Rix}
\affiliation{Max-Planck-Institut f\"ur Astronomie, K\"onigstuhl 17, D-69117, Heidelberg, Germany}

\begin{abstract}
Using \textit{Gaia}~DR3 astrometry and spectroscopy, we study two new substructures in the orbit-metallicity space of the inner Milky Way: \textit{Shakti} and \textit{Shiva}. They were identified as two confined, high-contrast overdensities in the $(L_z, E)$ distribution of bright ($G<16$) and metal-poor ($-2.5<\rm{[M/H]}<-1.0$) stars. Both have stellar masses of $M_\star \gtrsim 10^7M_\odot$, and are distributed on prograde orbits inside the Solar circle in the Galaxy. Both structures have an orbit-space distribution that points towards an \textit{accreted} origin, however, their abundance patterns -- from APOGEE -- are such that are conventionally attributed to an \textit{in situ} population. These seemingly contradictory diagnostics could be reconciled if we interpret the abundances [Mg/Fe], [Al/Fe], [Mg/Mn] \textit{vs.} [Fe/H] distribution of their member stars merely as a sign of rapid enrichment. This would then suggest one of two scenarios. Either these prograde substructures were created by some form of resonant orbit trapping of the field stars by the rotating bar; a plausible scenario proposed by \cite{Dillamore_2023}. Or, \textit{Shakti} and \textit{Shiva} were proto-galactic fragments that formed stars rapidly and coalesced early, akin to the constituents of the \textit{Poor Old Heart} of the Milky Way; just less deep in the Galactic potential and still discernible in orbit space.
\end{abstract}
%\keywords{Galaxy: halo  --  Galaxy: structure -- Galaxy: kinematics and dynamics -- chemical abundances -- surveys}

%%%%%%%%%%%%%%%%%%%%%%%%%%%%%%%%%%%%%%%%%%%%%%%%%%%%%%%%%%%%%%%%%%
%%%%%%%%%%%%%%%%% BODY OF PAPER %%%%%%%%%%%%%%%%%%%%%%%%%%%%%%%%%%
%%%%%%%%%%%%%%%%%%%%%%%%%%%%%%%%%%%%%%%%%%%%%%%%%%%%%%%%%%%%%%%%%%
\section{Introduction}\label{sec:Introduction}

One focus of modern astrophysics is to understand how massive galaxies, such as our Milky Way, formed and evolved. To address this broad theme with observations of the present epoch Universe, it has proven powerful to take the Milky Way itself as a galaxy model organism for which to map -- and then eventually understand -- the detailed age-orbit-abundance distribution of its stars. This constrains when and on which orbits its stars formed, how the stellar birth material became gradually enriched, and which dynamical processes -- e.g. mergers or secular evolution -- shaped the Galactic structure. The importance of various processes must depend both on the epoch and on the Galactocentric radius (or potential well depth) at which they occur.

\begin{figure*}
\begin{center}
\vspace{-0.20cm}
\includegraphics[width=\hsize]{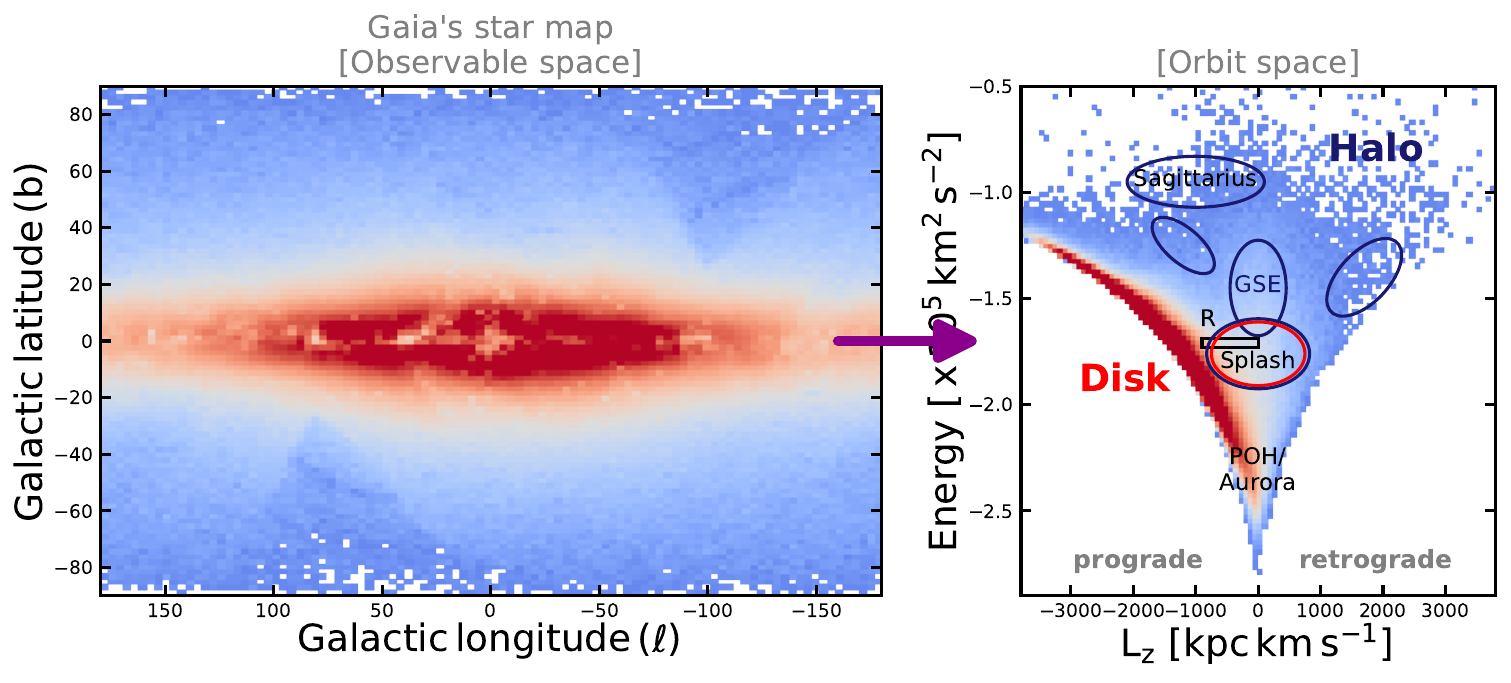}
\end{center}
\vspace{-0.6cm}
\caption{Schematic diagram showing the distribution of the Milky Way stars in the Galactic coordinates $(\ell, b)$ and in the orbit-space ($L_z, E$). This is based on the \Gaia\ data (see Section~\ref{sec:data}). The ``Disk'' highlights the \ins\ population (which comprises metal-rich stars with [M/H]$\simgt-1.0$) and the ``Halo'' highlights the \accreted\ population (which mostly comprises metal-poor stars [M/H]$\simlt-1.0$). Locations of some of the substructures are also highlighted, including \Sagittarius\ \citep{Ibata1994, Malhan_2022_etal}, \GSE\ \citep{Belokurov_2018, Helmi_2020, Malhan_2022_etal}, \Splash\ \citep{Belokurov_2020, Naidu_2020}, \cite{Dillamore_2023}'s ridge-like component (the rectangular box labelled as ``R'') and \Aurora/\POH\ \citep{Belokurov_2022, Rix_2022}; the location and extent of the drawn contours are only approximate.}
\label{fig:Fig1}
\end{figure*}

One of the most striking observations of our Galaxy is how dramatically age, metallicity, and orbits of stars are connected (e.g., \citealt{Bland-Hawthorn_2016}). The stellar body is dominated by a rotation-dominated, disky distribution (e.g., \citealt{Rix_2013, Helmi_2020}) that shows a bimodal density distribution in the [Fe/H]-[$\alpha$/Fe] plane \citep[e.g.][]{Bensby2014}, where both the kinematics and the spatial structure depend strongly on both [Fe/H] and [$\alpha$/Fe] \citep[e.g.][]{Bovy_2012}. Basically all these stars with disk-like kinematics have [M/H]$\simgt-1.0$ (e.g., \citealt{Mackereth_2019}). The central portions of our Galaxy ($D_{\rm gal}\simlt 4$~kpc) are dominated by a bar \citep{BlitzSpergel1991} and a central spheroid, dominated by random motions and a wide range of metallicities \citep[e.g.][]{Ness_2013,Belokurov_2022,Rix_2022}.  The arguably smallest component by stellar mass, but the largest by spatial extent (to $D_{\rm gal} \gtrsim 100$~kpc) is the stellar halo with [M/H]$\simlt-1.0$. At least beyond $D_{\rm gal}\sim 5$~kpc, it is predominately composed of unbound `debris' of tidally disrupted satellite galaxies \citep{Ibata1994, Helmi_1999, Bell_2008,Newberg_2009_Cetus, Belokurov_2018, Helmi_2018, Myeong2019, Koppelman2019, Yuan2020, Naidu_2020, Necib_2020, Malhan_2022, Tenachi_2022}.

Parsing the overall stellar orbit--abundance distribution into an extensive set of individual components has been a long tradition, as each of them may represent a discrete event or regime in our Galaxy's formation history. In particular, it has become established to categorise each component or substructure as either born from gas already within the Milky Way's dominant potential well (dubbed \ins) or born within a (at the time) distinct subhalo (or satellite galaxy)  that was later \accreted\ into the Milky Way. 

Among old ($\ge 8$~Gyrs) stellar populations in the Milky Way, the prototypical \ins\ population is the $\alpha$-enhanced or thick disk. This inference is based on the disk-like kinematics that reflects the settling of the stars' birth gas in the Milky Way's main potential well \citep[e.g.][]{Birnboim_2003,Stern_2021}, starting $t_{\rm age} \sim 12.5\Gyr$ ago \citep{Xiang_2022}. When looking at these stars in the $(L_z, E)$ plane of orbital parameters, their distribution follows the line of circular, in-plane orbits (see Fig.~\ref{fig:Fig1}), though not exactly owing to their substantive velocity dispersion. And this assessment of an \ins\ origin is based on the abundance patterns: these stars are enhanced in [Mg/Fe] and [Mg/Mn] and show a rapid rise of [Al/Fe] with [Fe/H]. All of these are taken as signatures of rapid enrichment that presumably requires high densities and a deep potential well \citep{Hawkins_2016,Belokurov_2022}.

Among the stellar halo components, the {\it Gaia-Sausage-Enceladus} (henceforth \GSE) population \citep{Helmi_2018,Belokurov_2018} exemplifies an \accreted\  population (see Fig.~\ref{fig:Fig1}; right panel). The very low angular momentum and large apocenters of their orbits cannot be readily explained by an \ins\ origin, and the low-$\alpha$ abundance patterns resemble those of the `surviving' dwarfs satellites of the Milky Way and point toward slow enrichment in a more diffuse potential well \citep{Wyse_2016, Hasselquist_2021}. 

Yet, when looking in detail at the very early phases of our Galaxy's evolution (say $\gtrsim 11-12\Gyr$ ago), this \ins~vs. \textit{accreted} distinction may become practically ambiguous and conceptually subtle. For instance, there is the ``Splash'' population (see Fig.~\ref{fig:Fig1}), identified by \cite{Bonaca_2017, Di_Matteo2019, Belokurov_2020} that has been termed the ``metal-rich in-situ'' halo: its orbits are radial (hence halo-like, favouring and \accreted\ origin), but the population is metal-rich with [M/H]$\simgt-1.0$ and $\alpha$-enhanced, pointing towards an \ins\ scenario. But both properties can be understood if its population arose from  stars that were kicked out of the early, nascent \ins\ disk into radial orbits during an early collision with some massive merger.

Recent studies of ancient stars in the inner Galaxy \citep{Kruijssen_2019, Belokurov_2022, Rix_2022, Horta_2022, Conroy2023} have identified a dynamically hot population of tightly-bound stars, namely the \textit{Poor Old Heart} of the Milky Way (or Aurora), henceforth \POH\ (see Fig.~\ref{fig:Fig1}). Most of these stars show the chemical signatures of rapid enrichment (i.e., these stars are enhanced in [Mg/Fe] and [Mg/Mn] and show a rapid rise of [Al/Fe] with [Fe/H]) that are conventionally attributed to an \ins\ origin. Simulations of galaxy formation (e.g., \citealt{Renaud_2021}) suggest that this component originated presumably from the coalescence of several proto-Galactic fragments that were of comparable, though not equal, mass. Whether at such early phases, it is useful to attribute the term \ins~ for only one of these presumed protogalactic fragments \citep{Belokurov_2022} is not clear \citep[cf.][]{Rix_2022,Conroy2023}. Nonetheless, the abundance pattern of these stars may serve as a reminder that they only imply rapid enrichment (which may require high star formation rates, deep potential wells and high densities), rather than as a direct indicator of the dynamical pre-history.

But also the inference path from a particular orbit distribution to, say, an \accreted\ origin may not be straightforward or unique. For instance, as mentioned above, the ``Splash'' population has halo-like orbits and appear to favour an  \accreted\ origin; but, it actually comprises metal-rich stars with [M/H]$\simgt-1.0$ that suggest it was the earliest part of the $\alpha$-enhanced disk whose stars were dramatically stirred during an early, quite massive merger. Likewise, the ``R'' feature indicated in Fig.~\ref{fig:Fig1}, which represents ridge-like overdensity of stars, may lend itself to the interpretation as an \accreted\ population. But \cite{Dillamore_2023} showed that this feature comprises both metal-rich and metal-poor stars and their orbits lie somewhere between disk- and halo-like. Given that these stars' azimuthal orbit frequency coincides with that of the Galactic bar, this orbit-space substructure can be very plausibly attributed to the trapping of the (inner) halo stars in bar resonances.

In this study, our primary goal is to examine the (inner) Milky Way's ($L_z, E$)--[M/H] space, using the recent \Gaia~DR3--based XGBoost dataset, \citep[][A23]{Rix_2022, Andrae_2023}, to look for previously unrecognised substructures. In doing so, we identified two orbit-abundance space overdensities that we call \Shakti\ and \Shiva, and show these possess very intriguing properties that send ``mixed messages'' about their \ins\ or \accreted\ origin. To this end, Section~\ref{sec:data} describes the data used. Section~\ref{sec:orbits} details our method to compute the orbital parameters of the stars. In section~\ref{sec:orbit-abundance_space}, we analyze the ($L_z, E$) distribution of stars in different slices of [M/H], thereby identifying \Shakti\ and \Shiva. Section~\ref{sec:Shakti} and Section~\ref{sec:Shiva} present the characterization of these stellar populations. In Section~\ref{sec:origin}, we lay out possible origin scenarios for these two populations. We finally conclude in Section~\ref{sec:Conclusion}.

\begin{figure*}
\begin{center}
\vspace{-0.20cm}
\includegraphics[width=0.90\hsize]{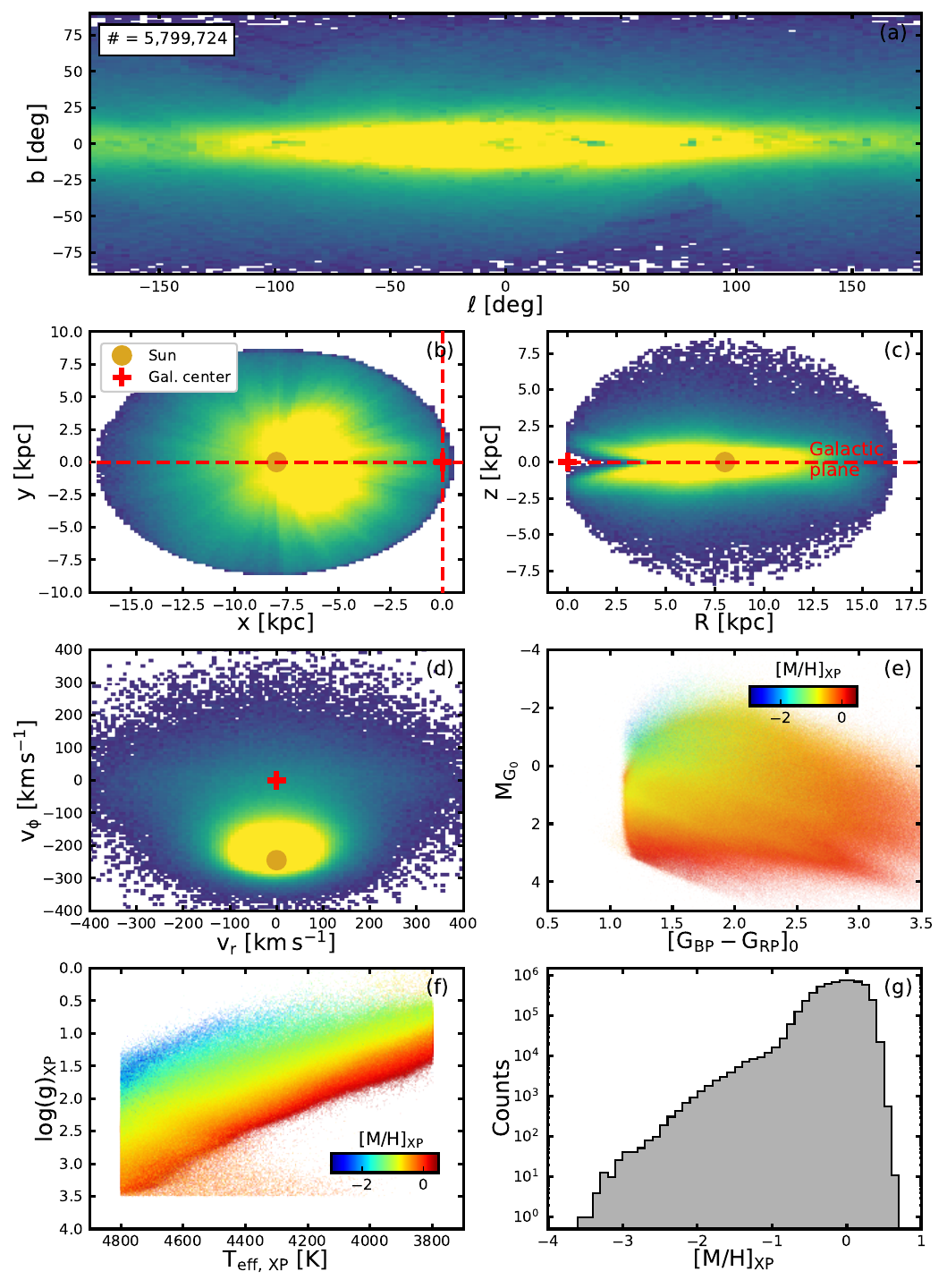}
\end{center}
\vspace{-0.5cm}
\caption{Overview of the overall dataset used in this analysis, without any kinematic or [M/H]$_{\rm XP}$ cuts. {\bf Panel (a):} Distribution of stars in the Galactic coordinates. The legend on the top-left mentions the total number of stars in our dataset. {\bf Panel (b):} Distribution of stars in the Galactocentric $x - y$ plane. In this Cartesian system, the Galactic center (denoted by ``+'') lies at the origin and the Sun (denoted by a yellow dot) is at $(x, y, z) = (-8.122, 0, 0)$~kpc. {\bf Panel (c):} Same as panel $b$, but in the Galactic $R-z$ plane. {\bf Panel (d):} Velocity behaviour of the sources in spherical polar coordinates, namely radial $v_r$ and azimuthal $v_\phi$. {\bf Panel (e):} Color-magnitude distribution (CMD), where each star is coloured by its $\rm{[M/H]_{\rm XP}}$ value. Note that the y-axis is the absolute magnitude. {\bf Panel (f):} Effective temperature ($T_{\rm eff, XP}$) vs surface gravity ($\rm{log(g)_{\rm XP}}$). {\bf Panel (g):} Metallicity distribution function (MDF).}
\label{fig:Fig2}
\end{figure*}
\section{Data}\label{sec:data}

To explore the Milky Way's stellar ($L_z, E$)$-$[M/H] space, we require for each star [M/H] information and a complete 6D phase-space measurement, i.e., its sky position ($\alpha, \delta$), parallax ($\varpi$, which can be inverted to obtain the heliocentric distance), proper motions ($\mu^{*}_\alpha \equiv \mu_\alpha \rm{cos}\delta, \mu_\delta$) and line-of-sight velocity ($v_{\rm los}$). 

\subsection{XGBoost dataset}

For this, one of the most extensive and well-vetted data sets is the recent \Gaia~DR3--based XGBoost catalogue \citep[][A23]{Rix_2022, Andrae_2023}. A23 provides data-driven XGboost estimates of stellar metallicity $\rm{[M/H]_{\rm XP}}$, surface gravity $\rm{log(g)_{\rm XP}}$ and effective temperature $T_{\rm eff, XP}$ for $175$ million sources. These  are based on their \Gaia~DR3 low-resolution BP/RP (or, XP) spectra, after being trained on stellar parameters from APOGEE that have been augmented by a set of very metal-poor stars. Although all of these stars have \Gaia~DR3's 5D astrometry, only brighter sources (with $G_{\rm RVS}\simlt 14$~mag) also have \Gaia\ RVS velocities. 
A23 combines one of the most extensive data set with 6D phase-space with [M/H] that are robust, precise, and vetted against a number of external catalogs, such as APOGEE~DR17 \citep{Abdurrouf_2021}, GALAH~DR3 \citep{Buder_2021}, LAMOST~DR7 \citep{Zhao_2012_LAMOST} and SEGUE \citep{Yanny2009}. In our analysis, we restrict ourselves to only those stars with $\texttt{phot\_g\_mean\_mag} < 16$, $3800$~K $\leq T_{\rm eff, XP} < 4800$~K and $\rm{log(g)_{\rm XP}}< 3.5$, because in this range XGBoost's [M/H] estimates are most robust and precise (with $\delta\rm{[M/H]}<0.1$). Even with these quality cuts, this A23 subset contains $\sim 30$ times more stars compared to APOGEE~DR17 and LAMOST~DR7, $\sim 70$ times more stars compared to GALAH~DR3 and $\sim5000$ times more stars than in SEGUE. 

We  apply additional quality to ensure good \Gaia\ astrometry \citep[similar to][]{Malhan_2022}: (1) \texttt{parallax\_over\_error}$\geq5$; this quality cut ensures that our resulting sample contains well-measured parallaxes with relative parallax error of $\leq20\%$. Effectively, it also ensures that $\varpi>0$ and that the estimated heliocentric distances $D_\odot (=1/\varpi)$ are physical. We correct for the parallax zero-point of each star as $\varpi_{\rm corrected} = \varpi_{\rm observed} - (-0.017)$ \citep{Lindegren_2021}. (2) \texttt{phot\_bp\_rp\_excess\_factor}$< 1.7$; this moderate cut removes stars with strong $G_{\rm BP}-G_{\rm RP}$ color excess that occurs in the very crowded regions of the Galaxy (for normal stars, \texttt{phot\_bp\_rp\_excess\_factor}$\approx 1$, \citealt{Arenou_2018}). (3) \texttt{ruwe}$<1.4$; this value has been prescribed as a quality cut for ``good'' astrometric solutions by \cite{Lindegren_2021}. This results in a sample of $5,915,741$ stars. 

Finally, we excise all the stars that lie within five tidal radii of the globular clusters (listed in \citealt{Harris_2010}) in the 2D projection space. This removes $116,017$ stars.

Our final `base' dataset comprises $n = 5,799,724$ stars, and its overview is provided in Fig.~\ref{fig:Fig2}. We correct for dust extinction using \cite{SFD_1998} maps and assuming the extinction ratios $A_G/A_V = 0.86117$, $A_{\rm BP}/A_V =1.06126$, $A_{\rm RP}/A_V = 0.64753$ (as listed on the web interface of the PARSEC isochrones, \citealt{Bressan_2012}). Henceforth, all magnitudes and colors refer to these extinction-corrected values. 

\subsection{Orbits}\label{sec:orbits}

To compute the orbital parameters of stars (such as  $L_z$, $E$, etc), we need to transform their heliocentric phase-space measurements into Galactocentric Cartesian coordinates. This is achieved using the Sun's Galactocentric location as $R_{\odot}=8.112\kpc$ \citep{Gravity2018} and the Sun's Galactocentric velocity as $(v_{x,\odot}, v_{y,\odot}, v_{z,\odot})=(12.9,245.6,7.78)\kms$ \citep{Drimmel_2018}. Next, the Galactic potential is set up using the \cite{McMillan2017} model, which is a static and axisymmetric model comprising a bulge, disk components and an NFW halo. With this potential, we compute the orbital parameters, using the \texttt{galpy} module \citep{Bovy2015}. 

\begin{figure*}
\begin{center}
%\vspace{-0.1cm}
\includegraphics[width=\hsize]{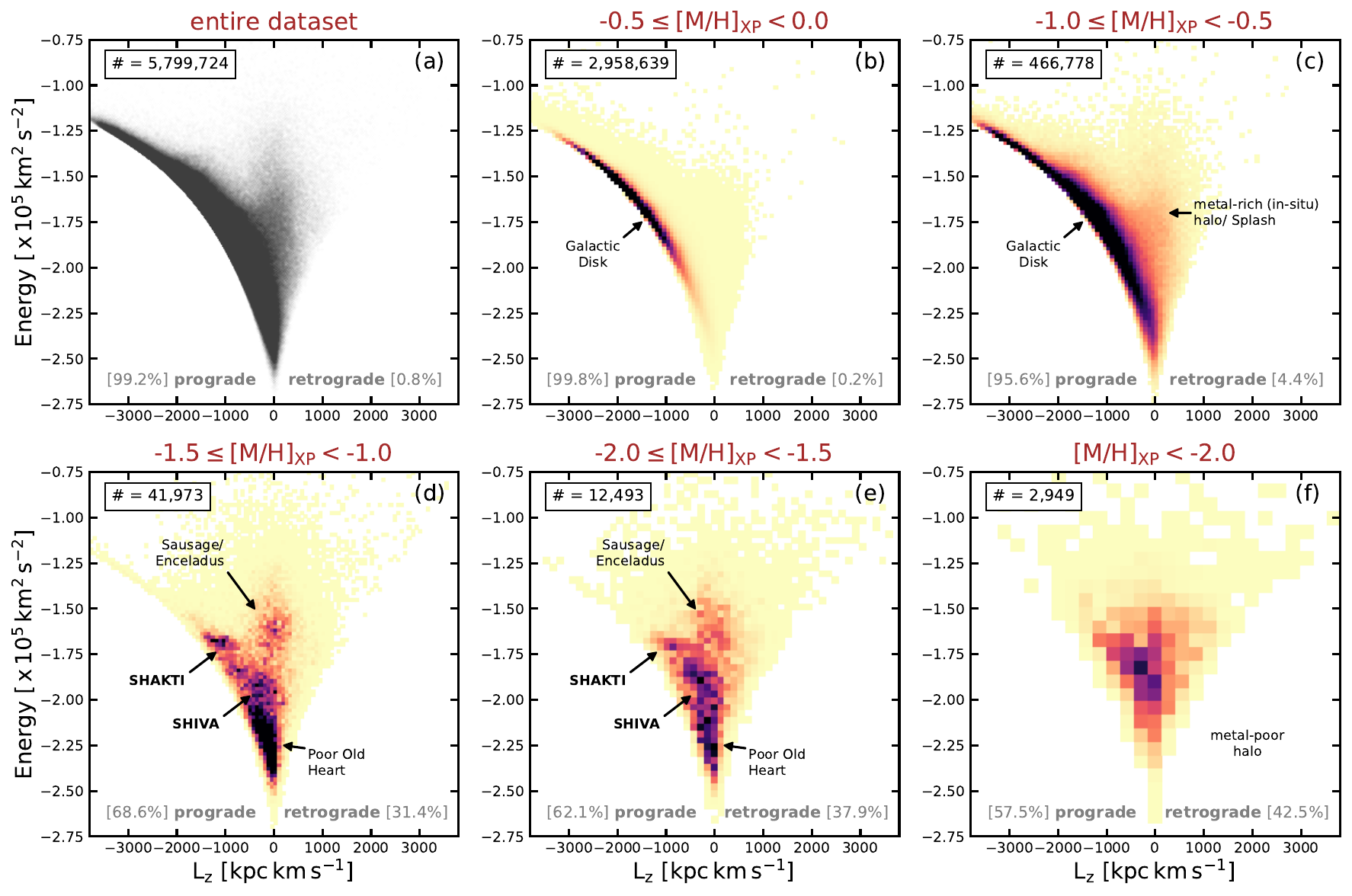}
\end{center}
\vspace{-0.5cm}
\caption{($L_z, E$) distribution of stars in different slices of metallicity $\rm{[M/H]_{XP}}$. {\bf Panel (a):} $(L_z, E)$ distribution of the entire dataset is shown as a scatter plot. {\bf Panel (b) -- (f):} $(L_z, E)$ distribution in different $\rm{[M/H]_{XP}}$ slices is shown as 2D histogram plots. $\rm{[M/H]_{XP}}$ decreases from panel (b) to (f). In every panel, the top text mentions the $\rm{[M/H]_{XP}}$ slice range and the top-left legend mentions the number of stars within this $\rm{[M/H]_{XP}}$ slice. In a given panel, the 2D histogram is of $(n_{\rm bin} \times n_{\rm bin})$ bins, where $n_{\rm bin} = \texttt{min}(100, 0.5\sqrt{n_{\rm data}})$ and $n_{\rm data}$ is the number of stars within that $\rm{[M/H]_{XP}}$ slice. The labelled substructures correspond to those that we visually identify as overdensities.}
\label{fig:Fig3}
\end{figure*}
\section{New stellar structures in $(L_z, E)-\rm{[M/H]}$ space:\\  Shakti and Shiva}\label{sec:orbit-abundance_space}

Fig.~\ref{fig:Fig3} shows the ($L_z, E$) orbit distribution of these stars in different $\rm{[M/H]_{XP}}$ slices, i.e. $n\bigl (L_z, E~|~\rm{[M/H]_{XP}}\bigr )$. Foremost, this Figure illustrates the known fact that the stellar orbit distribution depends dramatically on their metallicity, with the Galactic disk(s) dominating at $\rm{[M/H]_{XP}}>-1$ (see panels $b$ and $c$). It also shows clearly some of the well-established low-metallicity `sub-structures', such as the {\it metal-rich in-situ halo}/\Splash\ (\citealt{Bonaca_2017, Di_Matteo2019, Belokurov_2020}, visible in the panel $c$), \GSE\ merger debris (\citealt{Belokurov_2018, Helmi_2018}, visible in the panels $d$ and $e$) and \POH\ (\citealt{Belokurov_2022, Rix_2022}, visible in the panels $d$ and $e$).

Panels $d$ and $e$ of Fig.~\ref{fig:Fig3} (which together correspond to the metal-poor regime $-2.0\leq \rm{[M/H]_{XP}} < -1.0$) reveal the presence of three prominent substructures that are well confined in $(L_z, E)$ space. One of them with net $L_z \approx 0$ corresponds to the known \GSE\ structure. The other two prograde substructures at $(L_z, E)\sim (-1000 \kpc\kms, -1.7\times10^5\km2s2)$ and at $(L_z, E)\sim (-300 \kpc\kms, -1.9\times10^5\km2s2)$ have previously not been observed as such prominent overdensities or appreciated as possible substructures.  They are the focus of the present work, and for brevity, we denote them as \Shakti\ and \Shiva\ respectively\footnote{In Hindu philosophy, \Shiva\ -- \Shakti\ are the divine couple and their union (i.e., {\it shivashakti}) gives birth to the whole macrocosm. The goddess \Shakti\ represents the purest form of energy and the god \Shiva\ (one of the trinity gods) represents consciousness.}. The fact that these substructures are observed here for the first time with such high contrast, owes to the novel richness of the XGBoost dataset. 

In the remaining study, we characterize the stellar populations of \Shakti\ and \Shiva\ and aim to understand their nature and origin (i.e., \ins\ or \accreted). At this juncture, it is important to note that the $(L_z, E)$ location of \Shakti\ is similar to two previously identified substructures: the ridge-like component pointed out in \cite{Dillamore_2023}, who attribute them to halo stars that got spun up as they got trapped in the resonances of the Galactic bar (see our Fig.~\ref{fig:Fig1}); and also a sparse population of metal-poor disk stars (e.g., \citealt{Carollo_2019, Naidu_2020}). To what extent these substructures are identical, overlapping or connected with \Shakti\ is a question we address in Section~\ref{sec:origin}.

\begin{figure*}
\begin{center}
\vspace{-0.20cm}
\includegraphics[width=0.90\hsize]{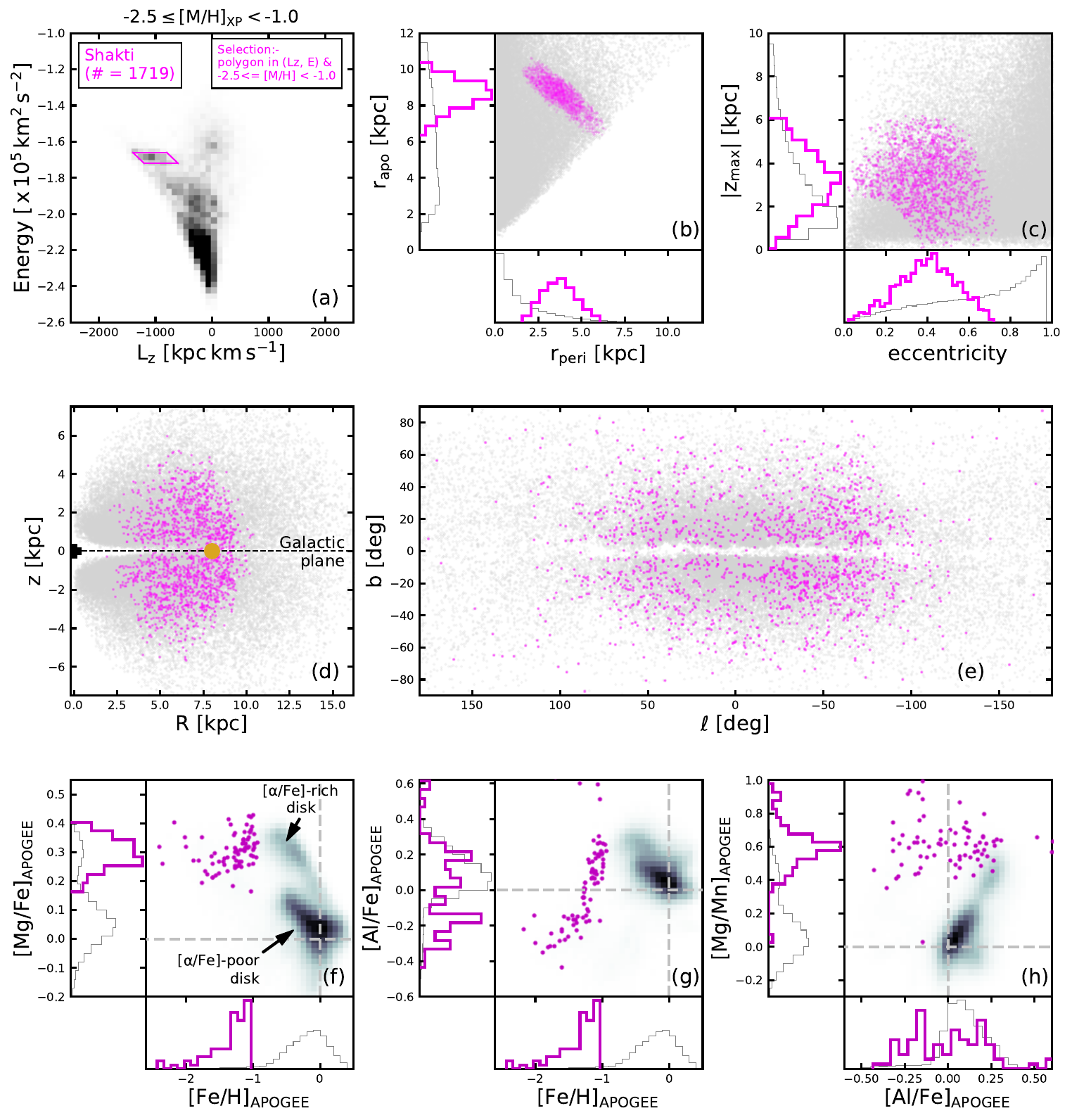}
\end{center}
\vspace{-0.4cm}
\caption{Distribution of the \Shakti\ stars in the chemo-dynamical space. {\bf Panel (a)}: \Shakti's selection function. {\bf Panel (b) - (c):} Distribution of stars in ($r_{\rm peri}$, $r_{\rm apo}$) and (eccentricity, $z_{\rm max}$). {\bf Panel (d) -- (e):} Distribution of stars in the Galactocentric $R-z$ plane and in the Galactic sky coordinates. {\bf Panel (f) -- (h):} Distribution of stars in the chemical space using the SDSS/APOGEE~DR17 measurements.}
\label{fig:Fig4}
\end{figure*}
\section{Characterization of Shakti}\label{sec:Shakti}

Below, Section~\ref{subsec:Shakti_stars} describes our procedure to select the \Shakti\ population, Section~\ref{subsec:Shakti_analysis} details its chemo-dynamical properties and in Section~\ref{subsec:Shakti_mass} we compute its stellar mass.

\subsection{Selection of \Shakti\ stars}\label{subsec:Shakti_stars}

We select \Shakti\ stars drawing on simple visual inspection of the $n\bigl (L_z, E~|~\rm{[M/H]_{XP}}\bigr )$ distribution, inspired by the high-contrast appearance among low-metallicity stars (see panel $d$ of Fig.~\ref{fig:Fig3}). Initially, we had tried for automated and systematic selection by employing some detection algorithms (namely, \texttt{ENLINK}, \citealt{Sharma_2009_ENLINK}, and Gaussian mixture model, \citealt{Pedregosa_2011}) and also experimented with different data sub-spaces (e.g., the 2D $(L_z, E)$ space, the 3D action $(J_R, J_\phi, J_z)$ space, the 4D $(\mathbf{J}, E)$ space, the 3D $(L_z, E, \rm{[M/H]_{XP}})$ space). We did not find that any of these served our specific purposes better than selection ``by hand''.

To select the \Shakti\ population, we use the base data and restrict ourselves in the metallicity range $-2.5 \leq \rm{[M/H]}_{XP} < -1.0$ and then retain only those stars that lie inside a 2D $(L_z, E)$ polygon described by four vertices as $[-1400, -1.66]$, $[-800,  -1.66]$, $[-600,  -1.72]$, $[-1200, -1.72]$; the units are in [$\kpc\kms, \times 10^5\km2s2$]. This polygon best describes this population. The [M/H] range is in concordance with the [M/H] slices in which we identify \Shakti\ in Fig.~\ref{fig:Fig3}, and we especially probe the low--[M/H] range to gather even the very metal--poor stars. This selection results in $N_{\rm sample}=1719$ stars.

\begin{table*}
\centering
\caption{Orbital and chemical properties of \Shakti\ and \Shiva\ stellar populations.}
\label{tab:Table1}
\begin{tabular}{c||c|c||c|c||}
\cmidrule{2-5}
&  \multicolumn{2}{c||}{\bf Shakti}  &  \multicolumn{2}{c||}{\bf Shiva} \\
\hline
\multicolumn{1}{|c||}{\bf Parameters} &  \multicolumn{1}{c|}{Median} & \multicolumn{1}{c||}{Spread} & \multicolumn{1}{c|}{Median} & \multicolumn{1}{c||}{Spread}\\
\hline

$r_{\rm peri}\,[\kpc]$ & $ 3.56 _{- 0.05 }^{+ 0.06 }$ & $ 2.03 _{- 0.08 }^{+ 0.06 }$ & $-$ & $-$\\
$r_{\rm apo}\,[\kpc]$ & $ 8.79 _{- 0.03 }^{+ 0.05 }$ & $ 1.71 _{- 0.07 }^{+ 0.06 }$ & $ 6.14 _{- 0.1 }^{+ 0.16 }$ & $ 2.41 _{- 0.13 }^{+ 0.12 }$\\
eccentricity & $ 0.41 _{- 0.01 }^{+ 0.01 }$ & $ 0.31 _{- 0.01 }^{+ 0.02 }$ & $-$ & $-$\\
$z_{\rm max}\,[\kpc]$ & $ 3.21 _{- 0.07 }^{+ 0.08 }$ & $ 2.57 _{- 0.1 }^{+ 0.11 }$ & $ 3.15 _{- 0.06 }^{+ 0.05 }$ & $ 1.82 _{- 0.15 }^{+ 0.15 }$\\
\hline
$\rm{[\alpha/Fe]_{APOGEE}}$ & $ 0.29 _{- 0.0 }^{+ 0.0 }$ & $ 0.08 _{- 0.01 }^{+ 0.01 }$ & $ 0.28 _{- 0.0 }^{+ 0.0 }$ & $ 0.09 _{- 0.01 }^{+ 0.01 }$\\
$\rm{[Mg/Fe]_{APOGEE}}$ & $ 0.3 _{- 0.0 }^{+ 0.01 }$ & $ 0.12 _{- 0.01 }^{+ 0.01 }$ & $ 0.3 _{- 0.01 }^{+ 0.01 }$ & $ 0.13 _{- 0.01 }^{+ 0.01 }$\\
$\rm{[Mg/Mn]_{APOGEE}}$ & $ 0.6 _{- 0.01 }^{+ 0.01 }$ & $ 0.24 _{- 0.02 }^{+ 0.02 }$ & $ 0.58 _{- 0.01 }^{+ 0.01 }$ & $ 0.24 _{- 0.02 }^{+ 0.02 }$\\

\hline
\end{tabular}
\tablecomments{The ``median'' represents the median of the distribution and the corresponding uncertainties represent their $16$ and $84$ percentiles. The ``spread'' is the inter--quartile difference between the $16$ and $84$ percentiles and the corresponding uncertainties represent their $16$ and $84$ percentiles.}
\end{table*}
\subsection{Orbital and chemical properties of \Shakti}\label{subsec:Shakti_analysis}

Fig.~\ref{fig:Fig4} shows the distribution of the \Shakti\ stars in the orbital space, spatial coordinates and chemical space. These distributions are quantified in Table~\ref{tab:Table1}, with uncertainties derived from bootstrap re-sampling. The basic properties can be summarized as follows.

\Shakti\ stars orbit the Milky Way within the Galactocentric distance range of $D_{\rm gal}\sim 2-10\kpc$ (Fig.~\ref{fig:Fig4}, panel $b$); they reach $z_{\rm max}\sim 1-6 \kpc$ away from the Galactic plane, and have moderate eccentricities of $e \sim 0.1-0.6$. Furthermore, these stars are well phase-mixed in the Galaxy (as shown in the $R-z$ and $\ell-b$ distributions).  

We now turn to understanding \Shakti's detailed abundance patterns. We initially had selected \Shakti~ members via $n\bigl (L_z, E~|~\rm{[M/H]_{XP}}\bigr )$. Now we aim to construct the detailed element ratios $p\bigl ( [X_1/X_2] | \rm{Shakti\,member}\bigr)$ that can serve as diagnostics to differentiate Galactic populations (e.g., \citealt{Hawkins_2015, Belokurov_2022}). Since Gaia DR3 cannot deliver these\footnote{\Gaia~DR3 RVS provides individual element abundances essentially only for stars with [M/H]$\simgt -1$ \citep{2023A&A...674A..29R}.}, we 
 cross--match the Gaia-identified \Shakti~ sample members with SDSS/APOGEE~DR17 \citep{Abdurrouf_2021,SDSS-DR18} to obtain [Fe/H], [Mg/Fe], [Al/Fe], [Mg/Mn]. In so doing, we find $n=81$ stars possessing [Fe/H] and [Mg/Fe] measurements. And of these, only $79$ stars possess [Al/Fe] measurements and $76$ stars possess [Mg/Mn] measurements.

\Shakti\ stars are metal--poor, with an MDF that  ranges from [Fe/H]$\approx-2.5$ to $-1.0$. The stars are all $[\alpha/$Fe]-enhanced, they show a tight relation in the  [Al/Fe]--[Fe/H] plane, and $50\%$ of stars have [Al/Fe]$>0$ (Fig.~\ref{fig:Fig4}, panel $f$, $g$ and $h$). Note that the [Fe/H]--[Al/Fe] relation is remarkably tight. The reason behind this is unclear, however, but it may be pointing towards a single, chemically simple origin for the \Shakti\ stars, rather than a mix of different populations. While CMD-based analyses are generally quite useful to study stellar populations, here this is not very helpful for two reaons: all \Shakti~ stars were chosen to be metal-poor, and hence by implication old, population of giant stars (see Section~\ref{sec:data} and panels $e$ and $f$ of Figure~\ref{fig:Fig2}). Second, for most members the parallaxes are not precise enough to enable meaninful age estimates.

How the above properties of \Shakti\ relate to its nature and origin is discussed below in Section~\ref{sec:origin} (together with that of \Shiva). 

\subsection{\Shakti's Stellar Mass}\label{subsec:Shakti_mass}

Obviously, the sample members that we attribute to \Shakti\ are a highly biased subset of that structure's stars, biased in luminosity and orbital phase.

If \Shakti\ reflected an accreted satellite, we could employ the much-used mass-metallicity relation for dwarf galaxies \citep{Kirby_2013}. Adopting  a mean [Fe/H]$_{\rm APOGEE}$ for \Shakti\ would imply $M_\star \simgt 10^7\msun$, as this mean metallicity resembles that of the known WLM dwarf galaxy, which has [Fe/H]$\approx -1.28$ and $M_\star \sim 1.7\times10^7\msun$ \citep{Zhang_2012, McConnachie_2012}. It is important to note that this [Fe/H]--$M_\star$ relation possesses significant scatter and also a redshift dependence (i.e., higher masses at higher redshifts at fixed [Fe/H], e.g., \citealt{Zahid_2014, Ma_2016}).

Alternatively, we can simply count \Shakti's stars, accounting for the fact that we only consider giants; and that we will not see its members at all orbital phases.  Indeed, \cite{Frankel_2023} suggested to exploit the fact that for phase-mixed structures the distribution in all three orbital angles $\Theta$ is expected to be uniform. If we can estimate the ``angle incompleteness'' $\lambda_{\rm sample}^{-1}$, then we can estimate \Shakti's total stellar mass, $M_\star^{\rm total}$ via 
\begin{equation}
M_\star^{\rm total} = \lambda_{\rm sample} \times N_{\rm sample}\times M_\star / {\rm giant~in~sample}\,
\end{equation}
where $N_{\rm sample}$ is the number of (giant) star sample members that we attribute to \Shakti\ (Fig.~\ref{fig:Fig4}a), $\lambda_{\rm sample}$ is the factor accounting for the sample incompleteness\footnote{By sample incompleteness we mean those stars that \Gaia\ (much like any other Galactic survey) fails to observe at greater distances beyond the solar neighbourhood, and also in the Galactic disk region due to the dust extinction (as can be seen from Fig.~\ref{fig:Fig2}).} and ``$M_\star / {\rm giant~in~sample}$'' is the stellar population mass per giant star (at a given absolute magnitude $M_{G_0} < M_{\rm G_0, max}$). These parameters are explained and computed in Appendix~\ref{appendix:stellar_mass}. In this way, we arrive at a stellar mass estimate for \Shakti~ of $M_\star^{\rm total}\sim 0.7\times 10^7~M_\odot$.

\begin{figure*}
\begin{center}
\vspace{-0.20cm}
\includegraphics[width=0.90\hsize]{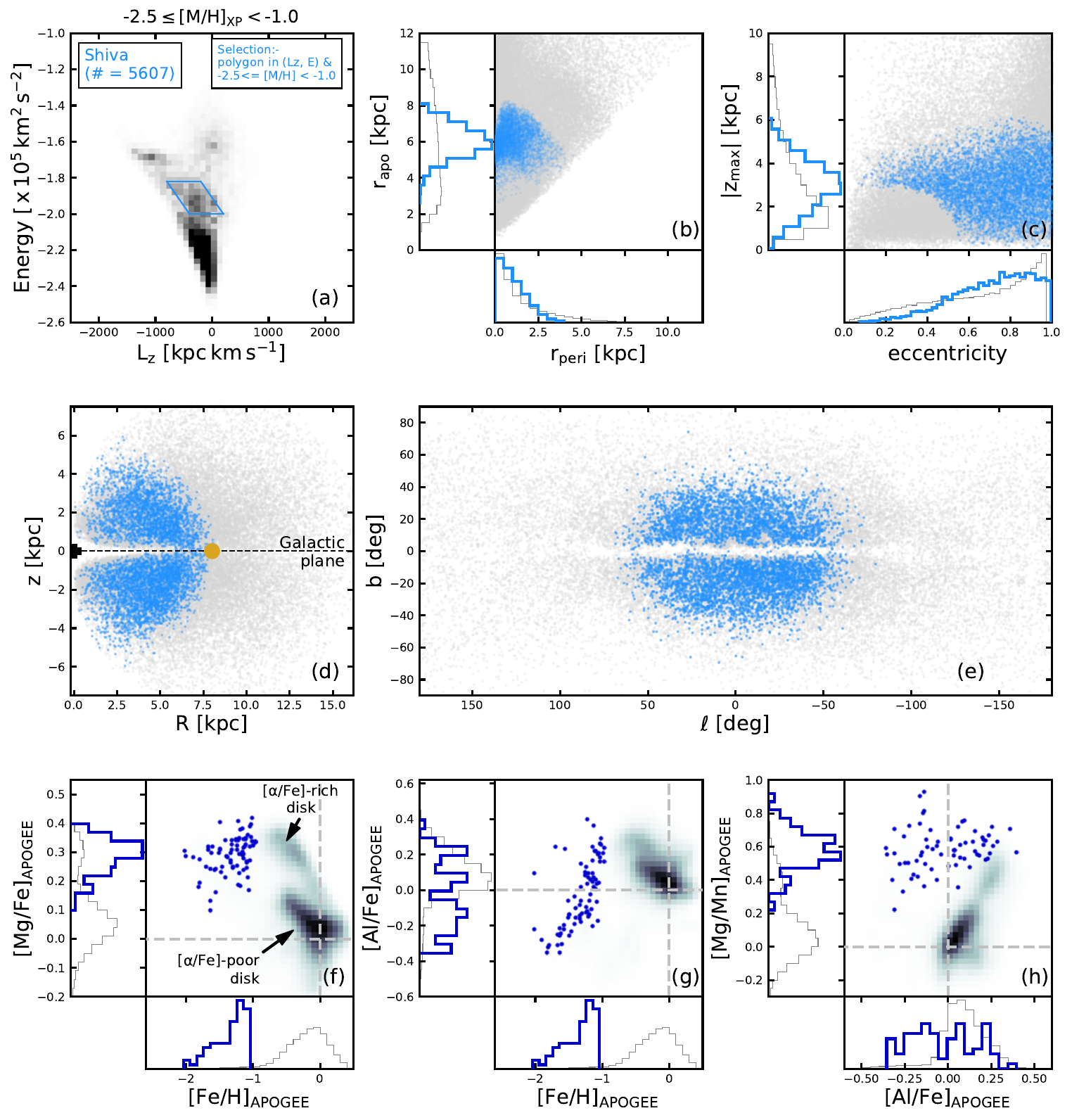}
\end{center}
\vspace{-0.4cm}
\caption{Chemo-dynamical properties of the \Shiva\ sub-structure. Same as Fig.~\ref{fig:Fig4}, but for the \Shiva\ stars.}
\label{fig:Fig5}
\end{figure*}
\section{Characterization of Shiva}\label{sec:Shiva}

For characterizing the \Shiva\ population, we follow the same approach as that described above for \Shakti\ (in Section~\ref{sec:Shakti}), except with slight modifications that are most suitable to study this substructure. 

The selection of \Shiva\ stars is again guided by the appearance of a prominent overdensity in panel $d$ of Fig.~\ref{fig:Fig3}. Specifically,  we restrict the full sample to the metallicity range $-2.5 \leq \rm{[M/H]}_{XP} < -1.0$ and then retain only those stars that lie inside the $(L_z, E)$ polygon described by the four vertices: $[-800, -1.82]$, $[-200, -1.82]$, $[200, -2.00]$, $[-400, -2.00]$; the units are in [$\kpc\kms, \times 10^5\km2s2$]. This selection results in $N_{\rm sample}=5607$ stars.

Fig.~\ref{fig:Fig5} shows the distribution of \Shiva\ stars in the orbital space, spatial coordinates and the chemical space, and these distributions are quantified in Table~\ref{tab:Table1}. These stars orbit the Milky Way within the Galactocentric distance of $D_{\rm gal} \sim 0-8\kpc$ (see $r_{\rm peri}-r_{\rm apo}$ distribution), with the amplitudes perpendicular to the Galactic plane ranging between $z_{\rm max} \sim 1-6\kpc$ and possess a moderately high range of eccentricities with $e \sim 0.2-1.0$. These stars are phase-mixed in the Galaxy (as shown in the $R-z$ and $\ell-b$ distributions). 

Next, we find \Shiva's detailed chemical abundance properties in an analogous fashion to that of \Shakti. In this case, with cross-matching of the Gaia-identified \Shiva\ sample members with APOGEE, we find $n=77$ stars possessing [Fe/H], [Mg/Fe] and [Al/Fe] measurements, and of these, only $68$ stars possess [Mg/Mn] measurements. We find it is relatively metal--poor, with a median of [Fe/H]$\approx-1.25$, it is quite rich in \textit{Mg}, \textit{Mn} and \textit{Al}, as listed in Table~\ref{tab:Table1}. Similarly to \Shakti , \Shiva 's population also has $50\%$ of stars with [Al/Fe]$>0$ and  shows a remarkably tight relation in [Fe/H] -- [Al/Fe]. 

To compute \Shiva's stellar mass, if we base an estimate on the mass-metallicity relation of dwarf galaxies \citep{Kirby_2013}, we arrive at $M_\star \simgt 10^7\msun$: e.g., similar to the WLM known WLM dwarf galaxy, which has [Fe/H]$\approx -1.28$ and $M_\star \sim 1.7\times10^7\msun$ \citep{Zhang_2012, McConnachie_2012}. On the other hand, if we count the \Shiva~member stars and correct them for (orbital) angle incompleteness and the sample's luminosity cut, we find $M_\star \sim 2.5 \times10^7\msun$, somewhat comparable to that of the \POH\ substructure \citep{Rix_2022} whose stars are distributed in the inner $D_{\rm gal}\sim3.5\kpc$ of the Galaxy and they possess [M/H]$<-1.0$. 

\begin{figure*}
\begin{center}
\vspace{-0.20cm}
\includegraphics[width=0.90\hsize]{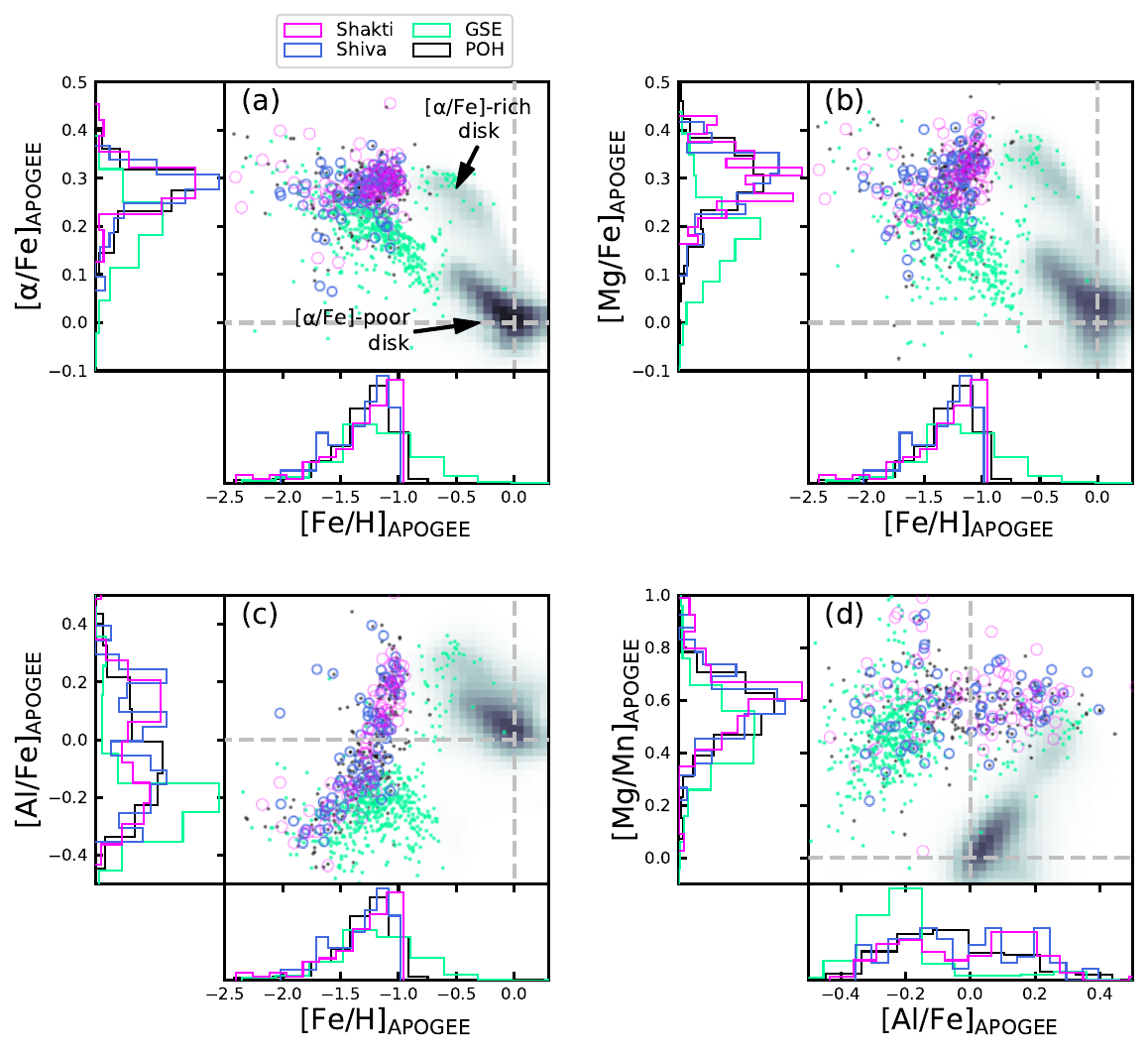}
\end{center}
\vspace{-0.4cm}
\caption{Comparing chemical abundances of different substructures.}
\label{fig:Fig6}
\end{figure*}
\section{On the ''Nature'' and origin of Shakti and Shiva}\label{sec:origin}

As we show below, \Shakti\ and \Shiva\ populations possess quite unconventional, intriguing properties that have not been observed previously in any substructure. This renders it quite challenging to understand the true nature and origin of these substructures. Below, we list down their properties and simultaneously interpret them to understand which of the following origin scenarios they possibly favour -- (a) \ins/disk origin (i.e., corresponding to the metal--rich disk population), (b) ``proto-Galactic'' origin (i.e., representing \Aurora/\POH--like metal--poor population, \citealt{Belokurov_2022, Rix_2022}), (c) dynamically ``heated'' disk origin (i.e., corresponding to the stars that were kicked out of the disk during an early collision with some massive merger, \citealt{Bonaca_2017, Di_Matteo2019}), (d) ``spun up'' halo population (i.e., representing those halo stars that got trapped in a ridge-like component in resonances with the Galactic bar, \citealt{Dillamore_2023}) or (e) \accreted\ origin (i.e., representing a population that was accreted inside a foreign galaxy, much like the \GSE). The following discussion is summarised in Table~\ref{tab:Table3}. 

\begin{table*}
\centering
\caption{Arguments on the nature and origin of \Shakti\ and \Shiva.}
\label{tab:Table3}
\begin{tabular}{| m{3.5cm} | m{7cm} | m{7cm} |}
\hline
{\bf Possible origin} & {\bf Shakti}  &  {\bf Shiva}  \\
\hline
                           &  \underline{supporting properties:}                    &                                                    \\
\ins/disk origin           &  $+$ High value of $|L_z|$                             &                                                    \\
(i.e. a part of the        &  $+$ possesses stars with [Al/Fe]$> 0$, implying       &                                                    \\
metal-rich disk            &      rapid enrichment inside a massive progenitor      &                                                    \\
population)                &                                                        &                                                    \\
                           &  \underline{seemingly inconsistent properties:}        & \underline{seemingly inconsistent properties:}     \\
                           &  $-$ observed as a prominent blob in $(L_z, E)$        & $-$ $(L_z, E)$ location different than the disk    \\
                           &  $-$ [M/H] is well below the oldest/most               & $-$ Too metal-poor to represent the disk population \\
                           &       metal-poor part of the disk                      &                                                     \\
                           &  $-$ orbits are not disk-like, but more halo-like      &                                                     \\
\hline
                           &  \underline{supporting properties:}                    & Same arguments as for \Shakti\ (on the left)       \\
``proto-Galactic'' origin  &  $+$ [M/H]-poor                                        &                                                    \\
(i.e., representing        &  $+$ chemical abundances similar to that of \POH.      &                                                    \\
\Aurora/\POH-like          &  $+$ possesses stars with [Al/Fe]$> 0$, implying       &                                                     \\
metal-poor population)     &      rapid enrichment inside a massive progenitor      &                                                    \\
                           &                                                        &                                                    \\
                           &  \underline{seemingly inconsistent properties:}        &                                                    \\
                           &  $-$ observed as a prominent blob in $(L_z, E)$        &                                                      \\
                           &  $-$ higher binding energy ($E$) than \Aurora/\POH\    &                                                      \\
\hline
                           &  \underline{supporting properties:}                    &                                                   \\
``Heated'' disk origin     &  $+$ $(L_z, E)$ only slightly away from that of disk   &                                                   \\
(i.e., a part of the disk  &  $+$ possesses stars with [Al/Fe]$> 0$, implying       &                                                    \\
population that was kicked &      rapid enrichment inside a massive progenitor      &                                                   \\
out during an early        &                                                        &                                                     \\
collision with a           &  \underline{seemingly inconsistent properties:}        &\underline{seemingly inconsistent properties:}        \\
massive merger)            &  $-$ observed as a prominent blob in $(L_z, E)$        &  $-$ Too metal-poor to represent the disk population  \\
                           &  $-$ [M/H] is well below the oldest/most               &                                                      \\
                           &      metal-poor part of the disk                       &                                                      \\
\hline
                           &  \underline{supporting properties:}                    &                                                       \\
``Spun up'' halo origin    &  $+$ $(L_z, E)$ location near the tip of               &                                                        \\
(i.e., a part of the halo  &      \cite{Dillamore_2023}'s ridge-like component      &                                                       \\
population that got        &                                                        &                                                         \\
trapped in a ridge-like    &  \underline{seemingly inconsistent properties:}        &  \underline{seemingly inconsistent properties:}        \\
component in resonances    &  $-$ abundances are very distinct from \GSE\           & $-$ Although prograde, but not near the $(L_z, E)$     \\
with the Galactic bar)     &  $-$ compact structure in the $(L_z, E)$ space,        &    location of  \cite{Dillamore_2023}'s primary ridge component \\
                           &      unlike \cite{Dillamore_2023}'s extended ridge     & $-$ abundances are very distinct from \GSE\           \\
                           &      located between $L_z \in [-1000, 0]\kms\kpc$      &                                                        \\ 
\hline
                           &  \underline{supporting properties:}                    & Same argument as for \Shakti\ (on the left)             \\
\accreted\ origin          &  $+$ observed as high-contrast blob in $(L_z, E)$      &                                                         \\
(i.e., represents a        &  $+$ orbits are halo-like                              &                                                          \\
population that was        &  $+$ [M/H]-poor                                        &                                                          \\
accreted inside a          &                                                        &                                                          \\
foreign galaxy             &  \underline{seemingly inconsistent properties:}        &                                                          \\
                           &  $-$ enhanced in higher-order elements and             &                                                          \\
                           &      possesses stars with [Al/Fe]$> 0$, implying       &                                                          \\
                           &      rapid enrichment inside a massive progenitor      &                                                         \\
\hline
\end{tabular}
\end{table*}

In doing so, we also require to compare the chemical abundances of \Shakti\ and \Shiva\ samples with those of \GSE\ (which represents a prototypical massive merger) and \POH. For this, we construct the \GSE\ sample using our base data and follow \cite{Limberg_2022}'s selection criteria: $|L_z| < 500\kpc\kms$ and $30\leq \sqrt{J_R} < 50$. This renders a sample of $8000$ stars. For \POH, we follow \cite{Rix_2022}'s recommendation: $\rm{[M/H]_{XP}}<-1.0$, $D_{\rm gal}<3.5\kpc$ and $r_{\rm apo}<8.0\kpc$. This renders a sample of $16000$ stars. These samples are cross-matched with the APOGEE dataset. 

\begin{itemize}[leftmargin=*]%[leftmargin=\parindent,align=left,labelwidth=\parindent,labelsep=0pt]

\item {\bf Both \Shakti\ and \Shiva\ populations are metal--poor, with their [Fe/H] well below the oldest/most-metal-poor part of the disk ($\rm{[Fe/H]^{\rm minimum}_{disk}}\approx-1.0$, \citealt{Mackereth_2019}). Secondly, they are enhanced in [$\alpha/$Fe], [Al/Fe] and [Mg/Mn], with a median [Al/Fe]$\approx0$ and $50\%$ of stars possessing [Al/Fe]$>0$.} These chemical properties are shown in Fig.~\ref{fig:Fig6} and are interesting for the following reasons.

First, \accreted\ metal--poor stars (with [Fe/H]$<-1.0$) are presumed to stem from mergers that are less massive and less dense than the Milky Way (e.g., \citealt{Helmi_2020}). Consequently, such \accreted\ stars should show chemical signatures of slower enrichment inside the low--mass progenitors \citep{Wyse_2016, Hawkins_2015}, which is reflected in their low [Al/Fe] ($\simlt-0.2$) and with practically all the stars possessing [Al/Fe]$<0$. This latter point is based on \cite{Hasselquist_2021} who analysed the five most massive dwarf galaxies accreted by the Milky Way  -- \LMC, \SMC, \Sagittarius, {\it Fornax} and \GSE; also see \cite{Horta_2022}.

Second, metal-poor stars whose [$\alpha/$Fe], [Al/Fe] and [Mg/Mn] abundances point towards rapid enrichment, most likely would have happened within a massive and dense progenitor. Recently, \cite{Belokurov_2022} analysed the APOGEE~DR17 dataset, and dubbed the entire population of metal--poor ([Fe/H]$\simlt-1.0$) and {\it Al}--rich ([Al/Fe]$>-0.1$) stars as \Aurora, attributing them to correspond to the Milky Way pre-disk \ins\ population that initially surrounded the Milky Way in a kinematically hot spheroidal configuration (and some of which later settled into the disk configuration, \citealt{Sestito_2020}), but at the present-day is sparsely distributed within the solar radius. This population may  overlap with the \POH\ identified by \cite{Rix_2022} as a dense central concentration of metal-poor, $\alpha$-enhanced stars. \cite{Rix_2022} suggest that these stars need not come from a single \ins\ progenitor,  but may come from a few distinct dense and massive progenitors that coalesced at high-redshift ($T\simgt12.5\Gyr$ ago) to form the proto-Galaxy. 

To examine whether \Shakti's and \Shiva's abundance patterns look similar to the \accreted--\GSE\ or to \Aurora/\POH\, we perform Kolmogorov-Smirnov (KS) test for the null hypothesis that the given two samples are drawn from the same distribution. \Shakti\ and \Shiva\ samples are statistically very similar to that of \POH, but very different than that of \GSE\ (the null hypothesis can be rejected at $>99\%$ level). Furthermore, these two substructures are even more enhanced in higher--order elements than \LMC\ and \SMC\ (for instance, compare our Fig.~\ref{fig:Fig6} with Figure~5 of \citealt{Hasselquist_2021}). Moreover, the abundances of \Shakti\ and \Shiva\ are also very similar to each other, suggesting a similar progenitor or formation history.

In sum, the fact that \Shakti\ and \Shiva\ are [Fe/H]--poor but show abundance patterns indicative of rapid enrichment, similar to \Aurora/\POH, favours a  ``proto-Galactic'' origin. Perhaps  \Shakti\ and \Shiva\ were very early distinct entities that were simply more rapidly enriched than \GSE\ (and also than \LMC\ and \SMC), perhaps they were more dense; at least at the time of accretion.

\item {\bf Both \Shakti\ and \Shiva\ populations are observed as high-contrast, tightly-bound overdensities in the $(L_z, E)$ space. Furthermore, both possess prograde motion, with \Shakti\ possessing a much higher $|L_z|$ value (similar to that of the disk at the given $E$) and \Shiva\ possessing a lower $|L_z|$ value. Moreover, they are more tightly bound in the Milky Way potential ($E$) than  \GSE\ but less tighly bound that the \POH.} These properties can be seen in Fig.~\ref{fig:Fig3}.

Observations of such prominent overdensities in the $(L_z, E)$ space have conventionally been considered unambiguous telltale signs of \accreted\ populations (e.g., \citealt{Helmi_2018, Naidu_2020, Yuan2020}), that were once born in a potential well that was distinct from the (current) Milky Way. 

While \Shakti 's  high angular momentum makes it tempting to connect it with  the  \ins/disk origin, this scenario seems implausible given its [Fe/H]$<-1.0$. Secondly, while its $(L_z, E)$ location is similar to that of \cite{Dillamore_2023}'s ridge--like component, its compact ``blob''-like appearance in $(L_z, E)$ space argues against them being a ridge of former halo stars that became trapped in resonances with the Galactic bar. The feature that  \cite{Dillamore_2023} identified had a lower contrast than \Shakti. If \Shakti\ were part of a resonance-driven ridge in orbit space, we would expect the chemical properties  to be similar to that of the halo stars, in particular to \GSE\ (because the majority of halo stars presumably originated from \GSE\ alone, \citealt{Helmi_2020}). But \Shakti\ is chemically different from \GSE, as shown by the abundance data.  Furthermore, we note that \Shakti's location in orbit space is broadly similar to a postulated  ``metal--poor disk'' (e.g., \citealt{Carollo_2019, Naidu_2020}). However, this feature is much more extended and of low contrast than \Shakti. 

If \Shakti\ and \Shiva\ were once distinct protogalactic fragments, they must have been incorporated $\gtrsim 12\Gyr$ ago; at least their star formation must have preceded the oldest and most metal-poor part of the disk \citep{Xiang_2022}. This is qualitatively supported by the idea of taking the binding energy ($E$) as a proxy for the accretion time, which places them between the \POH\ ($\simgt12.5\Gyr$, \citealt{Rix_2022}) and \GSE\ ($\simgt11\Gyr$, \citealt{Belokurov_2018, Helmi_2018}).

\item {\bf Both \Shakti\ and \Shiva\ populations are distributed within $D_{\rm gal}\simlt8-10\kpc$ of the Milky Way, appear phase-mixed, and possess moderately large $z_{\rm max}$ and eccentricity values.} These properties can be seen in Figures~\ref{fig:Fig4} and \ref{fig:Fig5}. This indicates that the stars lie on dynamically hot orbits, hotter than the oldest part of the ($\alpha$-enhanced) disk. This may be consistent with both the \accreted\ and the ``heated'' disk scenarios. However, the latter scenario seems unlikely, given that \Shakti\ and \Shiva\ are observed as high--contrast orbit--substructures (and not sparse distribution of stars, such as that observed in the \Splash\ substructure). Even their [M/H]--deficiency opposes any plausible connection with the disk.

\end{itemize}

\begin{figure*}
\begin{center}
\vspace{-0.20cm}
\includegraphics[width=\hsize]{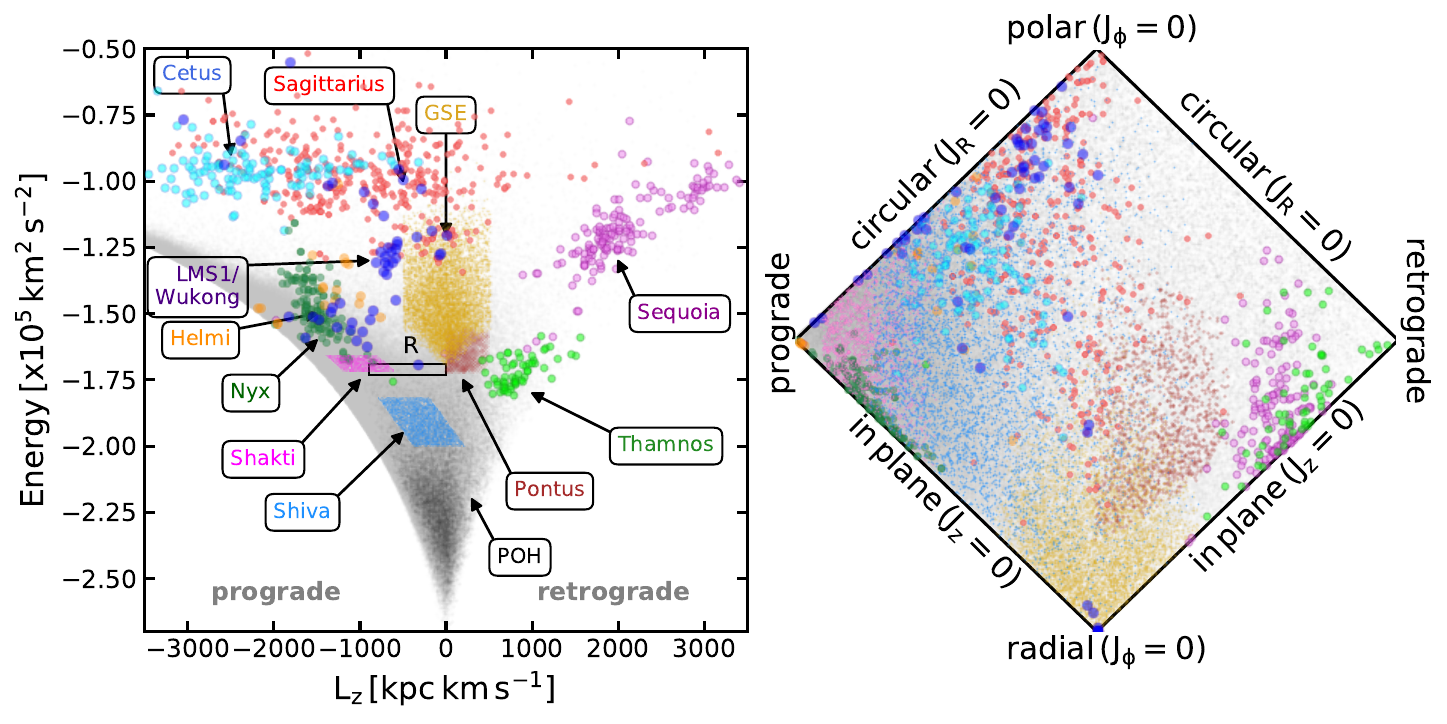}
\end{center}
\vspace{-0.4cm}
\caption{Milky Way's action-energy $({\bf J}, E)$ space showing stars from different substructures. Each data point corresponds to a \Gaia\ star. The left panel shows the $(L_z, E)$ distribution and the right panel shows the `projected action' space, where the horizontal axis is $J_{\phi}/J_{\rm tot}$ and the vertical axis is ($J_z$-$J_R$)/$J_{\rm tot}$ with $J_{\rm tot}=J_R+J_z+|J_\phi|$.}
\label{fig:Fig7}
\end{figure*}
In view of the above discussion, the following scenario appears most consistent with the totality of the observational evidence: \Shakti\ and \Shiva\ represent two of those early, massive progenitors that coalseced at high redshift (perhaps $T\simgt 12\Gyr$ ago), perhaps the last event form the proto-Galaxy, before disk formation commenced. These progenitors enriched more rapidly, and were likely even more dense and massive than \GSE\ (and also \LMC\ and \SMC), at least back then at the time of accretion. This conjecture is supported by the combination of their unconventional properties. 

To conclude the discussion, we ask whether \Shakti\ and \Shiva\ have any population of globular clusters dynamically ``associated'' with them? This is an interesting question because recent studies have shown that it is common for massive mergers to bring in populations of such objects (e.g., \citealt{Massari_2019, Myeong2019, Malhan_2022_etal}). To this end, it seems reasonable to link those Milky Way globular clusters that possess orbits and metallicities similar to these substructures. For the globular clusters, we obtain their orbits and metallicities from \cite{Malhan_2022_etal} and \cite{Harris_2010}, respectively. Next, we consider only those globular clusters that lie within the same $(L_z, E)$ polygon and [M/H] range as that described to select the substructure stars. \Shakti\ is found to be linked with only $n=1$ cluster (namely, NGC~6235) and \Shiva\ with $n=7$ clusters (namely, NGC~5986, NGC~6121/M~4, NGC~6218/M~12, NGC~6287, NGC~6333/M~9, NGC~6397,NGC~6809/M~55). 

\section{Conclusion and Discussion}\label{sec:Conclusion}

We have identified two new stellar structures in the Milky Way, seen as prominent, confined overdensities in the $(L_z, E)$ distribution of bright ($G<16$), metal--poor ($-2.5 < \rm{[M/H]} < -1.0$) stars. This was enabled by using the novel XGBoost dataset \citep{Rix_2022, Andrae_2023} that provides robust and precise metallicity estimates (based on \Gaia~DR3's BP/RP spectra) that can be combined with 6D phase-space measurements (from \Gaia~DR3 itself) for millions of stars; practically for the highest number of stars ever. More detailed chemical abundance patterns were enabled by  SDSS~DR17. 

\Shakti\ and \Shiva\ populations possess an unconventional combination of orbital and abundance properties that have not been observed previously. This is because it has become possible only recently to combine \Gaia\ and SDSS~DR17 datasets to disentangle different Galactic populations (e.g., based on the stellar orbits, [M/H] and [Al/Fe]), even in the inner Galaxy and at old ages, plausibly unearthing ancient ($T\simgt12.5\Gyr$ old) populations that formed the proto--Galaxy \citep{Belokurov_2022, Rix_2022}. There may be a preponderance of observational clues that argue for \Shakti\ and \Shiva\ representing two of early, massive ($M_\star\sim10^7\msun$) -- or at least dense -- fragments that coalesced at high redshifts (perhaps $z\simgt5$ or $T\simgt12\Gyr$ ago) to form the proto--Galaxy. They are less tightly bound than  \Aurora/\POH\ and therefore have a better chance of appearing as distinct overdensities in orbit-space.

However, there are important aspects in which  \Shakti\ and \Shiva\ match expectations for ancient stars that gradually picked up angular momentum by resonant orbit trapping caused by the Galactic bar; a scenario proposed recently by \citep{Dillamore_2023}. This scenario predicts ridges in the $E-L_z$ plane towards the prograde direction. And for the observationally inferred bar speed, these ridges should lie at orbital frequencies (or binding energies $E$) broadly similar to those of \Shakti\ and \Shiva\. Such orbital trapping should affect stars of all metallicities, and we see a fairly broad metallicity distribution in both substructures. However, quantitatively, this scenario does not match our observations: \Shakti's compact distribution in $L_z$ does not at all look like an extended ridge as observed in the \cite{Dillamore_2023} study. %it does not explain the compact distribution of \Shakti in $L_z$; \Shakti does not at all look like a ridge in our data
Nor does it explain the abundance pattern mismatch between \Shakti\ and \GSE, which is the presumed low-$L_z$ precursor in this scenario. Whether these inconsistencies, which argue against a spun-up-by-bar scenario, can be reconciled remains to be seen.

In closing, we put \Shakti\ and \Shiva\  in the dynamical context of the various other sub structures detected in the past. Fig.~\ref{fig:Fig7} shows different substructures of the Milky Way\footnote{The member stars of these different populations are obtained as follows. The {\it Sagittarius} stars are taken from \cite{Ibata_2020}, the {\it Cetus} population from \cite{Yuan_2022}, {\it LMS-1/Wukong} from \cite{Malhan_2021}, {\it Nyx} from \cite{Necib_2020, Wang_2022}, {\it Helmi} stream stars from \cite{Koppelman_2019}, {\it Pontus} from \cite{Malhan_2022}, {\it Thamnos} and {\it Sequoia} from \cite{Koppelman2019}. For \POH\, we follow \cite{Rix_2022}'s recommendation and select stars with $\rm{[M/H]_{XP}}<-1.0$, $D_{\rm gal}<3.0\kpc$ and $r_{\rm apo}<8.0\kpc$.}, most of which have been discovered using the \Gaia\ dataset, and their detailed chemical abundances are generally studied using the SDSS, LAMOST or GALAH surveys.  Excitingly, \Gaia's excellent all--sky 5D astrometry and metallicities are now being complemented by SDSS-V \citep{SDSS-V}, in particular also for \Shakti\ and \Shiva . In the near future, these spectroscopic efforts will be complemented by  measurements from the upcoming WEAVE \citep{Dalton_2014} and 4MOST \citep{deJong_2019} surveys that will map the northern and southern hemispheres, respectively. Furthermore, observations by the  Rubin Observatory (fka LSST, \citealt{Ivezic_2019}) will eventually provide a means to determine precise distances of the halo stars. These surveys together will provide 6D phase--space measurements and abundances of stars out to $\simgt50\kpc$, thereby allowing us to create a high--resolution chemo--dynamical map of different Galactic populations, giving us once in a lifetime opportunity to unravel the complete formation history of our Galaxy. 

\section*{Acknowledgements}

We thank the referee for helpful comments and sug-
gestions. We also thank Vasily A. Belokurov for enlightening and stimulating discussions on the topic. 

This work has made use of data from the European Space Agency (ESA) mission {\it Gaia} (\url{https://www.cosmos.esa.int/gaia}), processed by the {\it Gaia} Data Processing and Analysis Consortium (DPAC, \url{https://www.cosmos.esa.int/web/gaia/dpac/consortium}). Funding for the DPAC has been provided by national institutions, in particular the institutions participating in the {\it Gaia} Multilateral Agreement.

Funding for the Sloan Digital Sky Surveys IV and V has been provided by the Alfred P. Sloan Foundation, the Heising-Simons Foundation, the National Science Foundation, and the Participating Institutions. SDSS acknowledges support and resources from the Center for High-Performance Computing at the University of Utah. The SDSS web site is \url{www.sdss.org}.

SDSS is managed by the Astrophysical Research Consortium for the Participating Institutions of the SDSS Collaboration, including the Carnegie Institution for Science, Chilean National Time Allocation Committee (CNTAC) ratified researchers, the Gotham Participation Group, Harvard University, Heidelberg University, The Johns Hopkins University, L'Ecole polytechnique f\'{e}d\'{e}rale de Lausanne (EPFL), Leibniz-Institut f\"{u}r Astrophysik Potsdam (AIP), Max-Planck-Institut f\"{u}r Astronomie (MPIA Heidelberg), Max-Planck-Institut f\"{u}r Extraterrestrische Physik (MPE), Nanjing University, National Astronomical Observatories of China (NAOC), New Mexico State University, The Ohio State University, Pennsylvania State University, Smithsonian Astrophysical Observatory, Space Telescope Science Institute (STScI), the Stellar Astrophysics Participation Group, Universidad Nacional Aut\'{o}noma de M\'{e}xico, University of Arizona, University of Colorado Boulder, University of Illinois at Urbana-Champaign, University of Toronto, University of Utah, University of Virginia, Yale University, and Yunnan University.

\appendix

\begin{figure*}
\begin{center}
\vspace{-0.20cm}
\includegraphics[width=\hsize]{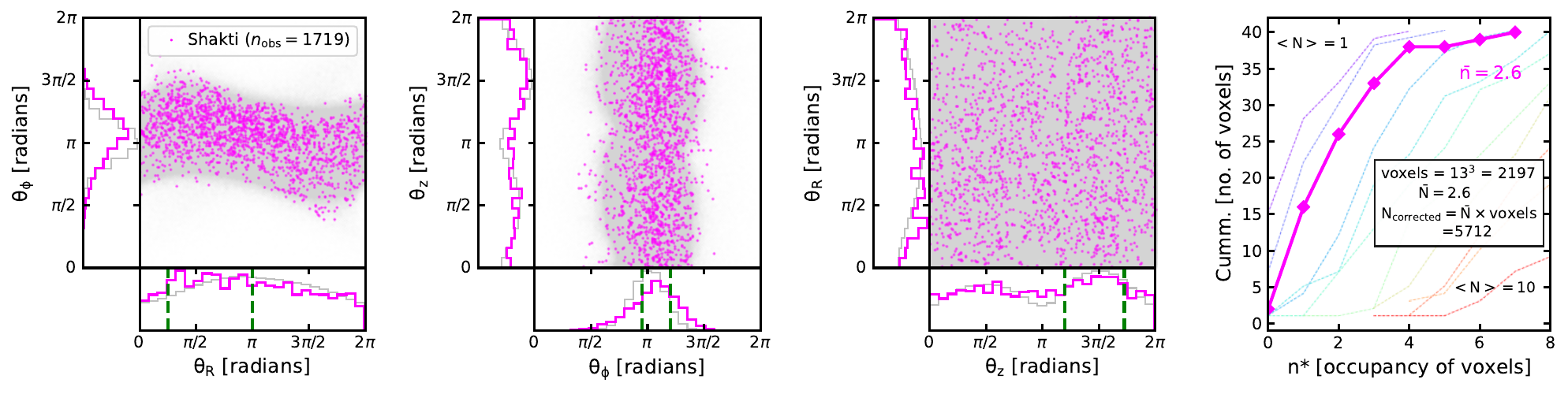}
\includegraphics[width=\hsize]{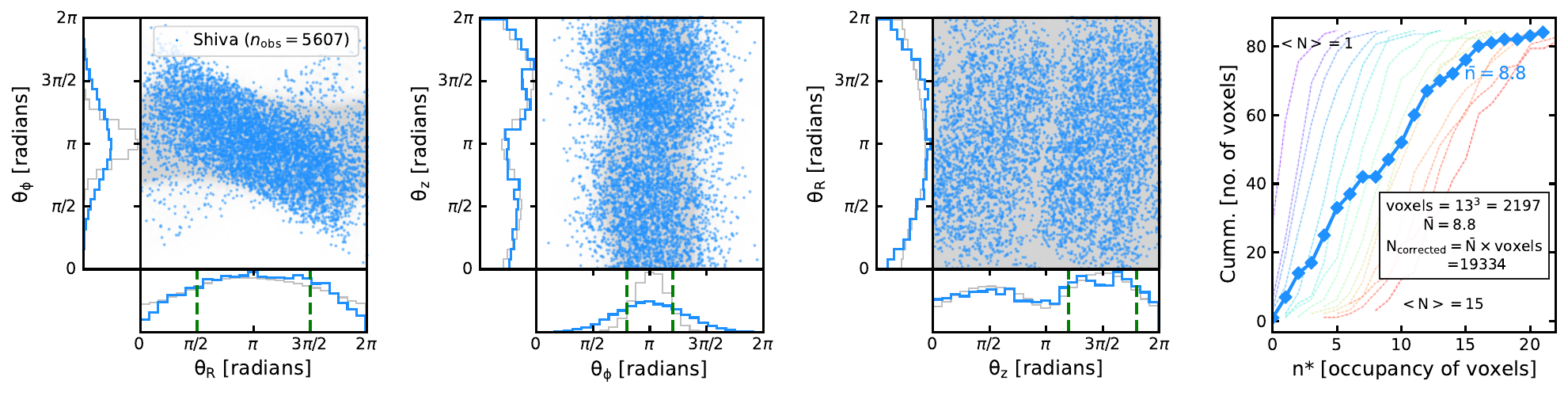}
\end{center}
\vspace{-0.4cm}
\caption{Accounting for the spatial selection function of \Shakti\ and \Shiva 's sample members. In each row, the first three panels show the distribution of the stars in the 3D space of orbital (action) angles, $\Theta  = (\theta_R, \theta_\phi, \theta_z)$ space. The overall sample (without any [M/H]$_{\rm XP}$ cut) is also plotted in the background. The last panel shows the estimate of the average occupancy number $\bar{n}$.}
\label{fig:Fig_appendix}
\end{figure*}
\section{Computing stellar mass of a phase-mixed population}\label{appendix:stellar_mass}

To compute the stellar mass of \Shakti\ and \Shiva, we use the following idea. We can simply count the member stars, accounting for the fact that we only possess giant stars in our sample (and the lower luminosity stars are missing) and that we observe member stars in a specific range of orbital phases (and not at all the phases). The latter occurs because \Gaia\ (much like any other Galactic survey) fails to observe stars at greater distances beyond the solar neighbourhood, and also in the Galactic disk region due to the dust extinction (as can be seen from Fig.~\ref{fig:Fig2}). In this regard, we note that \cite{Frankel_2023} suggests that for phase-mixed substructures (such as \Shakti\ and \Shiva), the distribution in all three orbital angles $\Theta = (\theta_R, \theta_\phi, \theta_z)$ is expected to be uniform. If we can estimate the ''incompleteness'' factor $\lambda_{\rm sample}^{-1}$ in the angle space, then we can estimate substructure's total stellar mass, $M_\star^{\rm total}$ via 
\begin{equation}\label{eq:appendix}
\begin{aligned}\
M_\star^{\rm total} &= \lambda_{\rm sample} \times N_{\rm sample}\times M_\star / {\rm giant~in~sample} \\
&= N_{\rm corrected}\times M_\star / {\rm giant~in~sample}.
\end{aligned}
\end{equation}
Here, $N_{\rm sample}$ is the number of (giant star) sample members that we attribute to the substructure (shown in Figures~\ref{fig:Fig4}a and \ref{fig:Fig5}a), $\lambda_{\rm sample}$ is the factor accounting for the sample incompleteness and ``$M_\star / {\rm giant~in~sample}$'' is the stellar population mass per giant star (at a given absolute magnitude $M_{G_0} < M_{\rm G_0, max}$). It is this parameter whose estimate relies on the above assumption. ``$M_\star / {\rm giant~in~sample}$'' is the stellar population mass per giant star (at the absolute magnitude $M_{G_0} < M_{\rm G_0, max}$). Below, we detail our procedure to compute the stellar mass.

\subsection{Shakti}

First, we compute the parameter $\lambda_{\rm sample}$. The fact that the sample is incomplete can be realised from Fig.~\ref{fig:Fig_appendix} which shows that the distribution of stars in the $\Theta$ space is not uniform. This is because we can only observe giant stars in our quadrant of the Galaxy. However, based on our above assumption for a phase-mixed population, the expectation is that this $\Theta$ distribution would have been uniform, had we been able to observe all the \Shakti\ stars in the entire Galaxy. Therefore, we require to estimate the mean occupation number $\bar{N}$, which is the average number of stars in a given $\Theta$ voxel. For this, we manually select the $\Theta$ sub-space where the distribution appears to be most complete (as shown in Fig.~\ref{fig:Fig_appendix}): $\rm{\theta_R}\in [ 0.8 , 3.1 ]$, $\rm{\theta_\phi}\in [ 3.0 , 3.8 ]$, $\rm{\theta_z}\in [ 3.8 , 5.4 ]$ and set the voxel bin size as $\rm{\theta_{bin}}= 0.5$ radians. The resulting cumulative distribution of \Shakti\ stars is shown in Fig.~\ref{fig:Fig_appendix}. Upon comparing this curve with the Poisson curves, we find the average stellar occupancy in a given voxel as $\bar{N} = 2.6$. This value needs to be multiplied with the total number of voxels, i.e., $2\pi/\rm{\theta_{bin}} = 13^3 = 2197$, which gives the total number of stars after correcting for sample incompleteness as $N_{\\rm corrected}= 5712$. This implies $\lambda_{\rm sample} = N_{\rm corrected}/N_{\rm observed} =  5712/1719= 3.3$.

Second, we compute the parameter ``$M_\star / {\rm giant~in~sample}$'' following the same approach described in \cite{Rix_2022}. For this, we use the globular cluster NGC~362, which serves as a template for an old, [M/H]$=-1.25$ population. Beyond NGC~362's half-mass radius of $1.2'$ \citep{Baumgardt_2018}, we find about $95$ giants with $M_{G_0} < 0$ (this magnitude limit is set by the faint limit of the \Shakti\ stars after the above $\Theta$ sub-space selection). Given NGC~362's total mass of $252,000\msun$ \citep{Baumgardt_2018}, this implies one giant (at $M_{G_0}<0$) per $1300\msun$ of the total stellar population mass. 

Finally, the above $\lambda_{\rm sample} (=3.3)$ and $M_\star / {\rm giant~in~sample} (=1300\msun)$ values are substituted in equation~\ref{eq:appendix} to obtain the stellar mass as $M_\star^{\rm total} \approx 0.7 \times10^7\msun$.

\subsection{Shiva}

To compute \Shiva's stellar mass, we follow the same procedure as above, but with slight modifications that are most suited for this stellar population. These include changes in the range of $(\theta_R, \theta_\phi, \theta_z)$, where we find \Shiva's distribution to be most complete: $\rm{\theta_R}\in [ 1.6 , 4.7 ]$, $\rm{\theta_\phi}\in [ 2.5 , 3.8 ]$, $\rm{\theta_z}\in [ 3.8 , 5.7 ]$. In this case, we find $\bar{N} = 8.8$, which in turn gives $N_{\\rm corrected}= 19334 $. This implies $\lambda_{\rm sample} = N_{\rm corrected}/N_{\rm observed} =  19334/5607= 3.4$. Next, ``$M_\star / {\rm giant~in~sample}$'' is computed similarly as above using NGC~362. In this way, we estimate for \Shiva\, $M_\star^{\rm total} \approx 2.5 \times10^7\msun$.

\bibliography{ref1}
\bibliographystyle{aasjournal}
%%%%%%%%%%%%%%%%%%%%%%%%%%%%%%%%%%%%%%%%%%%%%%%%%%

\end{document}